\documentclass[twocolumn,showpacs,preprintnumbers,amsmath,amssymb,aps]{revtex4}

\usepackage{graphicx}
\usepackage{amssymb}
\usepackage{epstopdf}
\usepackage{bm}

\newcommand\eqnn[2]{\begin{equation}#2\label{#1}\end{equation}}

\begin{document}

\title{ Covariant energy-momentum and an uncertainty principle for general relativity}

\author{ F. I. Cooperstock\thanks{{\tt cooperst@uvic.ca}}\\Department of Physics and Astronomy\\University of Victoria \\P.O. Box 3055, Victoria, B.C. V8W 3P6 (Canada)\\and M. J. Dupre\thanks{{\tt mdupre@tulane.edu}} \\Department of Mathematics\\Tulane University\\New Orleans, LA 70118 (USA)\\}



%
%



\begin{abstract}

We introduce a naturally-defined totally invariant spacetime energy expression for general relativity incorporating the contribution from gravity. The extension links seamlessly to the action integral for the gravitational field. 
The demand that the general expression for arbitrary  
systems reduces to the Tolman integral in the case of stationary bounded distributions, leads to the matter-localized Ricci integral for energy-momentum in support of the energy localization hypothesis. The role of the observer is addressed and 
as an extension of the special relativistic case, the field of observers
comoving with the matter is seen to compute the intrinsic global energy of a
system. 
The new localized energy supports the Bonnor claim that the Szekeres collapsing dust solutions are energy-conserving.
It is suggested that in the extreme of strong gravity, the Heisenberg Uncertainty Principle be generalized in terms of spacetime energy-momentum.

\end{abstract}


\maketitle


\vspace{.3in}

\section{Introduction}

The unresolved problems of energy and the issue of its localization within general relativity were debated in the early years of the theory's formulation.
Given their fundamental importance, it is understandable that they have 
remained subjects of considerable interest even up to recent times. By way of a brief history, in the 1920s, some authors including Einstein and Eddington held the view that while one could work usefully with energy in the traditional sense as a global concept in general relativity, no satisfactory meaning could be attached to its localization, a situation unprecedented in all the rest of physics. Their belief was based on the manner in which energy-momentum for the gravitational field was incorporated into general relativity. Rather than having a \textit{bona fide} energy-momentum tensor $T_i^k$ \footnote{Latin indices range from 0 to 3 and Greek indices range from 1 to 3.  We use units in which G=c=1.}
in general relativity to incorporate energy-momentum as in the rest of physics, these authors relied upon an energy-momentum pseudotensor $t_i^k$, first introduced by Einstein, to play the equivalent role for gravity.
Unlike true tensors, this pseudotensor could be made to vanish at any pre-assigned point by an appropriate transformation of coordinates, rendering its status rather nebulous. The pseudotensor was introduced in order to incorporate a global energy and momentum into general relativity, a necessary exercise, it was felt, because gravity had not lent itself to inclusion in the energy-momentum tensor $T_i^k$ as it had for all other fields. All fields other than the gravitational field incorporated themselves into the energy-momentum tensor, and global energy-momentum conservation followed naturally through the vanishing of the ordinary divergence of the energy-momentum tensor,\footnote{Repeated indices imply summation.}
\eqnn{Eq100}{\partial{T_i^k}/\partial{x^k}= {T_i^k}_{,k}=0.}
By integrating (\ref{Eq100}) over a given 3-volume and applying the Gauss divergence theorem, one readily expresses the time-rate of change of energy and momentum in the given 3-volume as accounted for by the flux of energy and momentum over the bounding 2-surface of this 3-volume.

However, in general relativity, (\ref{Eq100}) no longer applies. Rather, it is replaced by the vanishing \textit{covariant} divergence of the energy-momentum tensor, viz.\footnote{A semi-colon denotes covariant differentiation.}
\eqnn{Eq101}{{T_i^k}_{;k}=0.}
Equation (\ref{Eq101}) is the local expression for energy-momentum conservation in general relativity. Through the covariant derivative, it brings the metric and hence gravity into the conservation statement.
However, with this new form, one can no longer write the integral conservation laws as was the case previously in special relativity without an essential modification, the introduction of the aforementioned pseudotensor
$t_i^k$. When this is done, (\ref{Eq101}) is re-expressed as a vanishing ordinary divergence,
\eqnn{Eq102}{
{(\sqrt{-g}(T_i^k+t_i^k))}_{,k}=0
}
where $g$ is the determinant of the metric tensor $g_{ik}$.

Through the years, other pseudotensors performing the same function as that of Einstein's pseudotensor were introduced but they all carried the stigma of being non-covariant objects. In addition, they were not symmetric and hence did not lend themselves to forming an angular momentum construct as does the symmetric energy-momentum tensor $T^{ik}$ of special relativity. Landau and Lifshitz \cite{LL} were able to produce a symmetric energy-momentum pseudotensor but their construct was not Lagrangian-based. Moreover, like all pseudotensors, their's suffered from a lack of general covariance and could be made to vanish at any pre-assigned point by an appropriate choice of coordinate system.

Over the years, the debate over energy-momentum in general relativity took some interesting turns.
Various authors including Bondi \cite{Bondi} argued that gravitational energy must be localizable while Misner and Sharp \cite{Misner} accepted that energy was localizable for cases of spherical symmetry but not otherwise. Related to the energy issue was that of the nature of gravitational radiation, waves of gravity presumably carrying energy off to infinity in analogy with electromagnetic radiation. For decades, many authors (including the present first author) had calculated energy fluxes via gravitational radiation using energy-momentum pseudotensors and it was widely believed that the process was placed on a secure footing by Bondi \cite{Bondi2} who developed a ``news function'' to describe the flux. However, it was later shown that the news function is related to the pseudotensor and hence it carries the drawbacks and limitations of the latter \cite{Madore}.

As a result of this history, the prevailing  majority opinion
would appear to be that the energy of the gravitational field itself is not localizable and for total energy of a system including gravity, one must at best, adopt a so-called ``quasi-local'' approach. This has largely grown out of the ADM \cite{ADM} analysis of energy and momentum for spatially asymptotically flat models which in turn generalizes the Komar mass and momentum for models admitting timelike Killing fields. The Komar mass and momentum are defined as 2-surface integrals which for boundaries, coincide with the Tolman integral via the Gauss Theorem. The proofs of the Positive Energy Theorem for the ADM mass and momentum under the assumption of the dominant energy condition, first provided by Schoen and Yau \cite{S&Y}  and the simpler spinor proof given by Witten \cite{Witten1} have led to several different definitions of quasi-local energy having various differing properties, advantages and disadvantages. (See \cite{Szab} and \cite{Wang} for discussions of recent developments in the quasi-local approach). Later authors have made the Witten spinor proof rigorous (e.g. \cite{P&T}) and generalized it to include the Bondi mass \cite{H&T} in the asymptotically flat case. In spirit, we can say that these approaches attempt to make sense of how an observer sees  
the energy density in regions away from his location as compared to the energy density at his own location event, 
and are thus formed using integration.  The Positive Energy Theorem itself has been a major factor in the acceptance of the quasi-local approach, but 
it must be noted that it depends on the assumption of the dominant energy condition.  However, recently, all the energy conditions of general relativity have come into question \cite{B&V}.
On the other hand, if we were to have an acceptable energy-momentum tensor for the total energy including the gravitational field, it would be natural for an observer wishing to evaluate the total energy in a remote region to require observers located at each event in the region to report the energy density they observe and then to integrate all the results.  Such an approach would not depend on any assumptions of energy conditions, but of course would depend on the choice of observers throughout the region. In what follows, we will return to the 
important role of the observers.

\section{Localized energy-momentum}

In a series of papers \cite{Coop}, the first author presented a new hypothesis that energy, including the contribution from gravity, was localized in the non-vanishing regions of the energy-momentum tensor $T_i^k$. This carries the significant consequence that gravity waves do not actually convey energy in the course of their propagation through the vacuum. (This is assuming that the waves really do exist and they must exist if the essential element of finite velocity of propagation of information in relativity holds).

Various reasons were advanced in support of the hypothesis. There was the work of Gurses and Gursey \cite{Gurses} showing that an exact gravitational plane wave, the simplest of all wave types, is of the Kerr-Schild class for which one can always find a coordinate system where \textit{all} the components of the gravitational energy-momentum pseudotensor vanish identically everywhere, not just at a pre-selected point, a much stronger result than the freedom to remove the pseudotensor locally. Thus, one would be hard-pressed to believe that energy is flowing in this case if the construct representing its energy can be transformed out of existence in one fell swoop. This is in contrast to the energy-momentum tensor in electromagnetism whose time-time component $T_0^0$ and time-space components $T_0^\alpha$ which represent respectively the energy density and Poynting vector of energy flux, are non-zero for an electromagnetic wave
and remain non-zero for all transformations of the coordinates. Various other reasons were advanced in support of the localization hypothesis but the goal of finding a tensorial construct for general relativity embodying such a localization, remained unrealized.

In a recent paper \cite{Dupre}, the second author provided arguments in favor of the Ricci tensor $R_{ik}$ as the essential tensor embodying energy-momentum in general relativity, i.e. for the inclusion of the contribution from gravity. Clearly this is in support of the localization hypothesis as the Ricci tensor vanishes in vacuum. The choice of the Ricci tensor has immediate appeal as it is generally covariant, a true tensor unlike the pseudotensor. As well, it is attractive from another standpoint: Bondi \cite{Bondi3} had noted that since the Riemann tensor characterizes the presence of spacetime curvature, i.e. the essence of gravity, in an invariant manner, it would be natural, he reasoned, if the Riemann tensor were to describe the energy-momentum including the contribution from gravity. Since the second-rank Ricci tensor is formed as a trace of the fourth-rank Riemann tensor, it carries in part the content of the latter and hence could be seen to embody at least the spirit of the Bondi idea.

In this paper, which develops from its first introduction as an essay \cite{CoopDupreEssay}, we approach the problem from a different direction. The most reliable identification that we can make with energy for a system that includes the contribution from gravity within the framework of general relativity, stems not from local measure but rather global measure. Tolman \cite{Tolman} was able to show that the total energy of an isolated stationary (i.e. having no explicit dependence on time) distribution of matter could be expressed as
\eqnn{Eq1}{
E=\frac{1}{4{\pi}}\int{R_0^0}\sqrt{-g}d^3x
}
where $g$, as above, is the determinant of the four-dimensional spacetime metric and $R_0^0$ is the mixed time-time component of the Ricci tensor. (Naturally the coordinate system that is chosen for the derivation is one for which the system shows no explicit time-dependence.) Its measure as the correct total energy is secured by its connection with the coefficient of the $1/r$ term in the asymptotic form of the $g_{00}$ component of the metric tensor. There is no ambiguity about this measure of total energy. Through the use of the Einstein field equations, this energy is most often expressed in terms of the components of the energy-momentum tensor as
\eqnn{Eq2}{
E=\int\left(T_0^0-T_1^1-T_2^2-T_3^3\right)\sqrt{-g}d^3x,
}
although its less-familiar form in (\ref{Eq1}) is of particular interest for us here.
Tolman used the pseudotensor to achieve this result and a more elegant approach that did not rely upon the pseudotensor was later applied in \cite{LL} which produced the same answer.

An immediate point to note is that for either expression (\ref{Eq1}) or (\ref{Eq2}), while the imbedded square root of the metric determinant is for the complete four-dimensional spacetime metric, the integral is over the three-space coordinate volume $d^3x$, an unnatural juxtaposition of elements. The integral would be one over the logically preferred proper volume if $d^3x$ were replaced by $d^4x$ to connect properly with $\sqrt{-g}$.

At this point, we consider what minimum modification we can make to (\ref{Eq1}) that would render an expression with wholly covariant elements for energy-momentum distributions, including the contribution from gravity. We seek the modification that would cover 
\textit{arbitrary} systems, systems that may have intrinsic time-dependence, while incorporating the demand that its energy component reduce to (\ref{Eq1}) \footnote{
That is apart from a necessary multiplicative factor of $t$, the time interval that it is being observed.
}
in the case of a bounded stationary distribution.
As a minimum, we must change to a \textit{spacetime} integral (i.e. replace $d^3x$ with $d^4x$) to incorporate a \textit{proper} volume element. As well, we must replace the $R_0^0$ component with the \textit{complete} Ricci tensor in the integrand to have covariant elements. We designate the resulting structure $^*E_i^k$, 
\eqnn{Eq3}{
^*E_i^k=\frac{1}{4{\pi}}\int{R_i^k}\sqrt{-g}d^4x.
}

This expands the original expression into ten independent components, since the Ricci tensor is symmetric.
In the case of bounded stationary systems, its $0-0$ component does give the correct answer for the energy, multiplied by the time over which the system is being analyzed. \textit{We propose that the integral (\ref{Eq3}) is the necessary generalization for energy-momentum measure in general relativity.}
We will refer to it as ``spacetime energy-momentum''.
In our view, it is the entirely natural generalization of the concept of energy-momentum for general relativity as we are, by necessity, engaged with a curved four-dimensional spacetime in general relativity. It is an expression of the inextricable link between space and time in general relativity. While the change is not dramatic for stationary systems as it simply multiplies the traditional value by the time interval being measured, it is of considerable interest and complexity for intrinsically dynamic systems. In the latter, the admixture with time carries through the Ricci tensor and in a non-trivial manner in the metric determinant. We are familiar with integrals of the form of (\ref{Eq3}) in field theory, integrals of tensors over proper spacetime volume.  Thus it should come as no surprise that general-relativistic energy-momentum should require such a structure. In fact, in $^*E_i^k$, there is a particularly valuable connection to the essence of general relativity: the trace of $^*E_i^k$, namely $^*E_k^k$,
\eqnn{Eq3a}{
^*E_k^k=\frac{1}{4{\pi}}\int{R_k^k}\sqrt{-g}d^4x
}
where $R_k^k$ is the Ricci scalar, $R$, is 
(apart from a constant multiplier),
the invariant field action integral $S_g$ for general relativity, 
\eqnn{3aa}{
S_g=-\frac{1}{16{\pi}}\int\sqrt{-g}Rd^4x=-\frac{1}{4}{^*}E_k^k.
}
This integral, with the addition of the action integral $S_m$ for the matter, has its first variation set to zero by the fundamental Principle of Least Action, 
\eqnn{Eq3b1}{
{\delta}S_g + {\delta}S_m=0
}
yielding the Einstein field equations \cite{LL}.
 Thus, our new spacetime energy-momentum structure 
is seamlessly interwoven with the essence of the theory itself.

While the $^*E_i^k$ integral is not a tensor and hence is not ``covariant'' in the sense that physicists use the term, 
a change in coordinates changes the integral in a uniquely determined manner. Indeed, given the curvature of spacetime in general relativity, 
we submit that this is all that could be expected for such a measure over an extended distribution. The most that researchers ever hoped to find in the way of covariance for the incorporation of gravity into the umbrella of energy-momentum 
in general relativity was a \textit{local} tensor but locally to this point, all that had been forthcoming to encompass gravity was with a pseudotensorial add-on 
to the energy-momentum tensor, i.e. $T^{ik} +t^{ik}$ which was not covariant. Its replacement with $R_i^k$, first proposed by the second author \cite{Dupre}, is covariant and moreover, given the curvature of spacetime, is the extent of covariance, apart from the possibly additional 
covariant elements within the extended integrals, that one can expect. As well, in what follows, we will use the new construct to produce a four-scalar energy for an extended system in analogy with the familiar procedure in special relativity.

As we introduced a new kind of energy, spacetime energy, in the same vein 
it is natural to extend the traditional angular momentum measure to include the contribution from gravity in a new ``spacetime angular momentum'' construct. We do so in the form  $^*M^{ikl}$ as \footnote{
This expression is even less ``covariant'' than the expression for the extended spacetime energy-momentum because it employs the non-tensorial spacetime coordinates (as opposed to their differentials) in the density.
}
\eqnn{9}{
^*M^{ikl}=\frac{1}{4{\pi}}\int\left(x^iR^{kl}-x^kR^{il}\right)\sqrt{-g}d^4x
}
where the spatial components $(x^1,x^2,x^3)$ of $x^i$ for angular momentum are necessarily the quasi-Cartesian coordinates $(x,y,z)$. As with the spacetime energy-momentum, it is localized within the matter distribution.

The $^*E_i^k$ structure brings to mind the integrals 
over spacelike hypersurfaces and over 2-surfaces which are also employed in the quasi-local approach.  For instance, using the Witten integral \cite{H&T} in the case where $R$ is a region foliated by asymptotically flat spacelike hypersurfaces $\Sigma_t$ for $t$ in the interval $J$ of real numbers, we can compare the ADM mass to our spacetime energy-
momentum.  Let $u^a=u^{AA'}$ be the unit timelike normal vector field to the foliation.  We take $\alpha_t$ to be an asymptotically constant solution of the Dirac-Weyl neutrino
equation $D_{AA'} \alpha^A=0$ on $\Sigma_t$ and form the null vector field $K_a=\alpha_A  \bar{\alpha_{A'}}.$  Here $D$ denotes the spacetime Dirac operator restricted to operate on
vector fields tangent to the foliation hypersurfaces. Let $K^a(t,\infty)$ be the asymptotically constant value of $K^a$ on $\Sigma_t$ and let $P_a^{ADM}(t)$ denote the ADM  energy-momentum of $\Sigma_t.$ Witten's technique of proof for the positive energy theorem shows that

\eqnn{Eq1md}{
8\pi P_a^{ADM}(t)K^a(t,\infty)=$$$$\int_{\Sigma_t}[-u^{AA'}(D_b\alpha_A)(D^b\bar{\alpha_{A'}})+8\pi T_{ab}u^aK^b ]d\Sigma_t
}
which gives
\eqnn{Eq2md}{
8\pi P_a^{ADM}(t)K^a(t,\infty) =$$$$\int_{\Sigma_t}[-u^{AA'}(D_b\alpha_A)(D^b\bar{\alpha_{A'}})+(R_{ab}-(1/2)Rg_{ab})u^aK^b ]d\Sigma_t
}
for each time $t.$  Integrating over time on both sides then yields
\eqnn{Eq3md}{
4\pi E[u^a,K_b]-8 \pi \int_J P_a^{ADM}(t)d\tau=$$$$
\int_R [u^{AA'}(D_b\alpha_A)(D^b\bar{\alpha_{A'}})+(1/2)Rg_{ab}u^aK^b ]\sqrt{g}d^4x.
}
  The first term in the last integral is always negative, so any scalar curvature or metric condition guaranteeing negativity of the second term would mean that with appropriate consideration of the constant coefficients here, that the spacetime energy-momentum does not exceed the proper time integral of the ADM energy-momentum.

Returning to the enlarged structure of spacetime energy-momentum in (\ref{Eq3}),
we recall that this integral satisfies the requirements in a minimal sense.
It is well to ask if there is scope for a more complicated expression in the case of an intrinsically dynamic system. We consider the feasibility of developing an extension while maintaining the demand for the construction being composed of purely covariant elements. Any such extension must reduce to the correct Tolman expression (apart from a multiplication by the time of observation) in the case of bounded stationary systems. Suppose that for the general case, the Ricci tensor were to be replaced by the Ricci tensor plus an additional tensor (or tensors) of second rank. However, in order to reduce to the correct form for stationarity, the generalization must have partial derivatives with respect to time in such a form so as to reduce to the expression in (\ref{Eq1}) when the metric is stationary.  However, for the maintenance of purely covariant elements, such derivatives must be covariant derivatives. While one could envisage an infinite number of such forms, consider the following examples for different add-on tensors $S$ of different ranks:
\eqnn{Eq4}{
 R^i_k + S^i_{;k},     R^i_k  + S^{ij}_{j;k},     R^i_k  + S^{imj}_{j;km},
}
where a semi-colon represents covariant differentiation.
Regardless of the chosen form with covariant derivatives, while the desired partial derivatives with respect to time appear as required, extra undesired terms due to spatial derivatives appear as well, terms which persist even in the case of stationary systems, which is unacceptable. Thus, (\ref{Eq3}) is the only permissible form for spacetime energy-momentum with covariant elements in generality. 

The consequences are immediate: since the Ricci tensor is non-zero only in the regions where the energy-momentum tensor is non-zero, gravitational waves, waves of propagating spacetime curvature, are not carriers of energy-momentum through the vacuum, in conformity with the localization hypothesis \cite{Coop}.

\section{The role of the observer}

At this point, we return to the important role of the observer in relation to energy-momentum.
Recall that in special relativity we express the mass/energy $m$ of a body with four-momentum $p^i$ and four-velocity $u_i$ as the inner product 

\eqnn{Eq4a}{
 m=p^iu_i.
}
However $u_i$ can take on a broader role in the inner product; it can be taken as the four-velocity of an arbitrary observer.
With $u^i$ chosen as the four-velocity of the body 
itself, the result is the rest mass $m$. The observer measures the inner product as $m$ if he is comoving with the body. However, we could choose $u^i$ to be the four-velocity of an observer whose speed is $v$ relative to the body at the instant of his intersection with the body. Then the product $p^iu_i$ gives this particular observer's perception of the energy 
and its value is not $m$. Rather, its value is $m\gamma$ where $\gamma$ is the relativistic factor $(1-v^2/c^2)^{-1/2}$. The crucial role of observer is the lesson of great familiarity for us in special relativity 
where the particularly significant role of the comoving observer comes into play. 

Clearly, to extend this approach from a point to a distribution of energy, logically a continuum of observers is called for. It would appear natural to express the spacetime energy of a system in general relativity, relative to a continuum field of observers having a corresponding field of four-velocities $u^i$, in the form of a four-scalar, as\footnote{
With greater generality, we can view (\ref{Eq5}) as a special case of
$^*E[u^i,v_k]=\frac{1}{4{\pi}}\int{R_i^k}u^iv_k\sqrt{-g}d^4x$,
a bilinear functional of the fields $(u^i,v_k)$.
} 
\eqnn{Eq5}{
^*E=\frac{1}{4{\pi}}\int{R_i^k}u^iu_k\sqrt{-g}d^4x.
}
Moreover, following our experience with special relativity where the choice of the comoving observer in (\ref{Eq4a}) gave us the intrinsic mass, we naturally choose the field of comoving observers in (\ref{Eq5}) to extract the intrinsic energy for an extended system in general relativity. Indeed any other choice would lead to the inclusion of the $\gamma$ factors that we discussed above, factors antithetical to our search for the intrinsic energy.  

In the same vein, we can express the observer-related spacetime linear momentum $^*P^\alpha$ as 
\eqnn{Eq3b}{
^*P^{\alpha}
=\frac{1}{4{\pi}}\int{R^{\alpha}_iu^i}\sqrt{-g}d^4x
}
where $\alpha =(1,2,3)$.
The domains of integration are for our choosing according to the physical requirements.

\section{Applications of the spacetime energy integral}

As a very simple example, consider a spherically symmetric ball of matter at rest with exterior vacuum beyond its radius $a$. The metric can be expressed in the simple form in spherical polar coordinates
\footnote{
In fact arbitrary time-dependent spherically-symmetric distributions can be expressed in this form with $\nu=\nu(r,t), \lambda=\lambda(r,t)$ \cite{LL}.
}
\eqnn{Eq1ap}{
ds^2= e^{\nu(r)}dt^2-e^{\lambda(r)}dr^2- r^2[d{\theta}^2+sin^2{\theta}d{\phi}^2].
}
A well-known exact solution of this form with constant density equation of state $\rho={\rho}_0=constant$ was determined by Schwarzschild  
and the explicit form of the functions $\nu$ and $\lambda$ can be found in \cite{Tolman}.

Since the body is at rest, the field of comoving observers are rest observers in this frame in this case, with four-velocities
\eqnn{Eq2ap}{
u^i= (u^0,0,0,0).
}
From (\ref{Eq1ap}), we have the non-zero contravariant and covariant observer four-velocity components
\eqnn{Eq3ap}{
u^0=e^{-\nu/2}, u_0= e^{\nu/2}.
} 
Hence, from (\ref{Eq5}),
\eqnn{Eq4ap}{
^*E=\frac{1}{4{\pi}}\int{R_0^0}u^0u_0\sqrt{-g}d^4x
}
which reduces to the Tolman integral times the amount of time observed.


In fact, also dynamic spherically symmetric spacetimes are
of particular interest because of their simplicity. This is because the exterior vacuum for such systems is uniquely the static Schwarzschild solution as demonstrated in Birkhoff's theorem. For dynamic spherically symmetric interiors matching to the exterior Schwarzschild solution, the resultant spacetime energy is a simply separable product of the usual mass times the time-span of its observation. A simple example drawn from spherical dust collapse illustrates the result.

As shown in \cite{LL} and developed further in \cite{Coop2}, the essential equations in comoving spherical polar coordinates $(\tau, R, \theta, \phi)$ are as follows: The metric is
\eqnn{Eq1d}{
	ds^2= d\tau^2 - e^{\lambda} dR^2
		- r^2(d\theta^2 +\sin^2\theta d\varphi^2)
}
where $\lambda$ and $r$ are functions of $R$ and $\tau$. For the case in which the dust has been released from rest at infinity in the infinitely distant past, the Einstein field equations yield
\eqnn{Eq14d}{
	e^\lambda = (r')^2
}
\eqnn{Eq15d}{
	\dot{r}^2= \frac{F(R)}{r}
}
\eqnn{Eq16d}{
	r = \left(\frac{9F}{4}\right)^{1/3}
		( R - \tau )^{2/3}
}
\eqnn{Eq17d}{
	8\pi\rho = \frac{F'}{r' r^2}
}
where $\rho$ is the mass density, $F(R)$ is a function of integration, a prime denotes partial differentiation with respect to $R$ and a dot indicates partial differentiation with respect to $\tau$.
From (\ref{Eq17d}), it is easy to show with a simple integration \cite{LL} that $M(R)$, the total mass including the contribution from gravity, within the comoving radial coordinate $R$ is
\eqnn{Eq18d}{
	M(R)= F(R)/2.
}
From the metric (\ref{Eq1d}), the square root of the complete determinant is
\eqnn{Eq10}{
\sqrt{-g}=e^{\lambda/2}{r^2}sin\theta.
}
The 4-volume element in the comoving frame is
\eqnn{Eq11}{
{d{\tau}}d^3x= {d\tau}dRd{\theta}d\varphi.
}
Since dust is stress-free, in the comoving frame the only non-vanishing component of the energy-momentum tensor is $T_0^0$ and hence, from the Einstein field equations,
\eqnn{Eq12}{
R_0^0=4{\pi}T_0^0.
}
From (\ref{Eq1d}), we see that as in the static example above, the non-vanishing components of the four-velocity of the comoving observers are $u^0=u_0=1$.  
Thus, from (\ref{Eq5}),(\ref{Eq12}),(\ref{Eq10}),(\ref{Eq14d}),(\ref{Eq17d}),(\ref{Eq18d}) and (\ref{Eq11}), we find the \textit{spacetime} energy up to the comoving radius $R$ and over a proper time interval $\tau$ is
\eqnn{Eq13}{
\int\sqrt{-g}T_0^0d^4x=M(R)\tau
}
as expected. In spite of intrinsic time-dependence, for this special case, the result is simply separable in terms of standard energy and time because there are no gravitational waves emitted by the spherically symmetric system.




Gravity waves carry information from their source and hence the question arises as to whether this is inconsistent with these waves not being energy carriers. We know of no reason in principle to preclude the transfer of information in the absence of energy. Indeed Bonnor \cite{Bonnor} had noted that Szekeres \cite{Szekeres} asymmetric collapsing dust spacetimes have time-varying quadrupole moments and so are presumably energy-emitting according to the old ideas yet since they were asymptotically matched \cite{Bonnor} to the energy-conserving Schwarzschild form, cannot be losing energy, they are ``radiationless'', to use Bonnor's descriptor.  Our results are fully consistent with this finding. We see this as follows:

The Szekeres ``quasi-spherical'' collapsing spacetime consists of dust with metric \cite{Bonnor}, \cite{Szekeres}
\eqnn{Eq103}{
ds^2= dt^2-e^{\lambda}dr^2-e^{\omega}(dy^2+dz^2)
}
where $\lambda$ and $\omega$ are functions of $r,y,z$ and $t$. Solutions found by Szekeres are 
\eqnn{Eq104}{
e^{\omega/2}= \Phi(r,t)/P(r,y,z), e^{\lambda/2}= [P/W(r)]\frac{{\partial}e^{\omega/2}}{{\partial}r}
}
\eqnn{Eq105}{
P=a(r)(y^2+z^2) +2f(r)y +2g(r)z +c(r)
}
\eqnn{Eq106}{
ac-f^2-g^2=1/4
}
\eqnn{Eq107}{
\int\sqrt{(W^2-1+S/\Phi)}d\Phi=t +H(r).
}
In the solution, $a, f, g, c, W, S$ and $H$ are functions of integration, of which five are independent \cite{Bonnor}.

Since the metric is in synchronous form, the time lines
\eqnn{Eq108}{
u^i={\delta}^i_0
}
are geodesics and hence track the dust particle trajectories. Thus the dust, with density $\rho$ and energy-momentum tensor
\eqnn{Eq109}{
T_i^k={\rho}u_iu^k
}
is comoving with the coordinate system \cite{LL}.
After substitutions \cite{Bonnor}, 
\eqnn{Eq110}{
8\pi\rho=\frac{PS_1-3SP_1}{{\Phi}^2(P{\Phi}_1-{\Phi}P_1)}
}
where the subscript $1$ denotes an $r$ partial derivative. As expected, the density, through the function $\Phi$, is time-dependent (the density increases with collapse as time elapses).
  
The spacetime energy, using the four-velocities $u^i=u_i=(1,0,0,0)$ in (\ref{Eq5}) is
\eqnn{Eq111}{
^*E=\frac{1}{4\pi}\int{R_i^k}u^iu_k\sqrt{-g}d^4x=\frac{1}{4{\pi}}\int{R_0^0}\sqrt{-g}d^4x
}
and since dust has zero pressure, this reduces to
\eqnn{Eq112}{
^*E=\frac{1}{4{\pi}}\int\rho\sqrt{-g}d^4x.
}
To evaluate $\sqrt{-g}$, we note from (\ref{Eq103}) that
\eqnn{Eq113}{
\sqrt{-g}=e^{\omega+{\lambda}/2}.
}
Thus, we see that the metric determinant in addition to the density is time-dependent.
Using (\ref{Eq113}) and substituting the values of $\lambda$ and $\omega$ from (\ref{Eq104}) and the value of $\rho$ from (\ref{Eq110}) into (\ref{Eq104}), we find that the terms with $\Phi$, which are the only time-dependent terms, cancel. Thus, the spacetime energy is simply $t$, the observation time, multiplied by a constant and hence the standard 3-space energy is a constant, in agreement with the Bonnor ``radiationless'' deduction \cite{Bonnor}. It is particularly interesting to witness the time-dependence of the matter via the density $\rho$ coupling with the time-dependence of the geometry via the metric in $\sqrt{-g}$ to render the energy being conserved.


\section{Uncertainty principle for general relativity}

There is a new pathway that opens from the extension of energy-momentum to spacetime. This is suggested by noting that 
 (\ref{Eq5}) is of the form energy times time,  
the combination which manifests itself quantum-mechanically as a minimal product of uncertainties in the Heisenberg Uncertainty Principle
\eqnn{Eq5a}{
{\Delta}E{\Delta}t{\ge}\hbar/2,
}
where $\Delta$ denotes uncertainty. 

If the gravitational contribution to energy is localized, it is most reasonable to assume that gravity must partake in the uncertainty. 
This directive is amplified, as discussed above, by the crucial role of the observer regarding energy measurement.  We are thus guided to a generalization of the Heisenberg Uncertainty Principle for energy in the form
\eqnn{Eq8}{
\Delta\frac{1}{4{\pi}}\int{R_i^k}u^iu_k\sqrt{-g}d^4x{\ge}\hbar/2.
}  
It should be kept in mind that it is only for the regime of very strong gravity that this extended Uncertainty Principle would present a demonstrable difference relative to the standard Heisenberg form. 

Similarly, we generalize the standard Heisenberg expressions for linear momentum  
\eqnn{Eq7}{
{\Delta}P_x{\Delta}x{\ge}\hbar/2, {\Delta}P_y{\Delta}y{\ge}\hbar/2, {\Delta}P_z{\Delta}z{\ge}\hbar/2,
}
to the form
\eqnn{Eq3c}{
\Delta\frac{1}{4{\pi}}\int{R^{\alpha}_ku^k}\sqrt{-g}d^4x{\ge}\hbar/2.
}
The explicit role of the observer is evident in (\ref{Eq8}) and (\ref{Eq3c}) through the presence of the observer four-velocity $u^i$. 
We submit that it is in the context of general relativity, where spacetime finds its necessarily unifying character, that these generalized Uncertainty Principles to include gravity, arise so naturally.



\section{Concluding comments}

It is somewhat ironical to consider that Einstein's original focus was on the Ricci tensor for his field equations. The attention shifted to the Einstein tensor which is the non-gravitational energy-momentum tensor multiplied by constants. Now we are returning to the Ricci tensor, only now as the embodiment of full energy-momentum. It is noteworthy in its simplicity.

Many years ago, the late J.~L. Synge, one of the most distinguished mathematical physicists of the 20th Century,  expressed to the first author his sentiment that the concept of energy-momentum is simply incompatible with general relativity. His view was influenced by the pseudotensorial constructs for energy-momentum, widely discussed during this period, which were an anathema to him. While Synge's view seemed radical at the time, from the present perspective, we see it as conveying an essential element of truth: the standard energy-momentum concept does not mesh with general relativity for dynamic systems. However, the extended concept of \textit{spacetime} energy-momentum would appear to fit naturally, enabling us to maintain the concept of energy-momentum in general relativity.

\vskip 0.125in

{\bf Acknowledgment}
The authors are grateful to a referee for valuable comments.

\end{document}